# Asymptotically Regular LDPC Codes with Linear Distance Growth and Thresholds Close to Capacity

Michael Lentmaier[†], David G. M. Mitchell[*], Gerhard P. Fettweis[†], and Daniel J. Costello, Jr.[*]
[†]Vodafone Chair Mobile Communications Systems, Dresden University of Technology, Dresden, Germany,
{michael.lentmaier, fettweis}@ifn.et.tu-dresden.de
[*]Dept. of Electrical Engineering, University of Notre Dame, Notre Dame, Indiana, USA,
{david.mitchell, dcostel1}@nd.edu

***Abstract*— Families of *asymptotically regular* LDPC block code ensembles can be formed by terminating $(J,K)$-regular protograph-based LDPC convolutional codes. By varying the termination length, we obtain a large selection of LDPC block code ensembles with varying code rates and substantially better iterative decoding thresholds than those of $(J,K)$-regular LDPC block code ensembles, despite the fact that the terminated ensembles are almost regular. Also, by means of an asymptotic weight enumerator analysis, we show that minimum distance grows linearly with block length for all of the ensembles in these families, i.e., the ensembles are asymptotically good. We find that, as the termination length increases, families of "asymptotically regular" codes with capacity approaching iterative decoding thresholds and declining minimum distance growth rates are obtained, allowing a code designer to trade-off between distance growth rate and threshold. Further, we show that the thresholds and the distance growth rates can be improved by carefully choosing the component protographs used in the code construction.**

## I. Introduction

Low-density parity-check (LDPC) convolutional codes [1], the convolutional counterparts to LDPC block codes [2], have been shown to be capable of achieving the same capacity-approaching performance as LDPC block codes with iterative message-passing decoding. $(J,K)$-regular LDPC block code ensembles, with constant variable and check node degrees, have minimum distance that grows linearly with block length for $J > 2$, i.e., they are *asymptotically good*; however, they also have comparitively poor iterative decoding thresholds. LDPC codes based on a *protograph* [3] (or *projected graph* [4]) form a subclass of multi-edge type codes that have been shown to have many desirable features, such as good iterative decoding thresholds and, for suitably-designed protographs, linear minimum distance growth (see, e.g., [5], [6]).

So-called *asymptotically regular* LDPC block code ensembles [7] are formed by terminating $(J,K)$-regular protograph-based LDPC convolutional codes. This construction method results in LDPC block code ensembles with substantially better thresholds than those of $(J,K)$-regular LDPC block code ensembles, despite the fact that the ensembles are almost regular (see, e.g., [7]). By means of an asymptotic weight enumerator analysis [8], we show that the asymptotically regular LDPC code ensembles in this family are also asymptotically good. We find that, as the termination factor $L$ increases, we obtain families of codes with capacity approaching iterative decoding thresholds and declining minimum distance growth rates, allowing a code designer to trade-off between distance growth rate and threshold. Further, we show that the structure of the convolutional code is crucial to both the thresholds and growth rates of the asymptotically regular families. By carefully choosing the component protographs that form the convolutional protograph, we show that both the iterative decoding threshold and the minimum distance growth rate of the ensemble can be improved. Moreover, by increasing the complexity (measured by the average node degree), we show that it is possible to significantly improve both the growth rates and the thresholds as the termination factor gets large.

## II. Analysis of protograph-based LDPC codes

A protograph is a small bipartite graph $B = (V, C, E)$ that connects a set of $n_v$ variable nodes $V = \{v_0, \ldots, v_{n_v-1}\}$ to a set of $n_c$ check nodes $C = \{c_0, \ldots, c_{n_c-1}\}$ by a set of edges $E$. The protograph can be represented by a parity-check or *base* biadjacency matrix $\mathbf{B}$, where $B_{x,y}$ is taken to be the number of edges connecting variable node $v_y$ to check node $c_x$. Figure 1 shows an example of an irregular protograph.

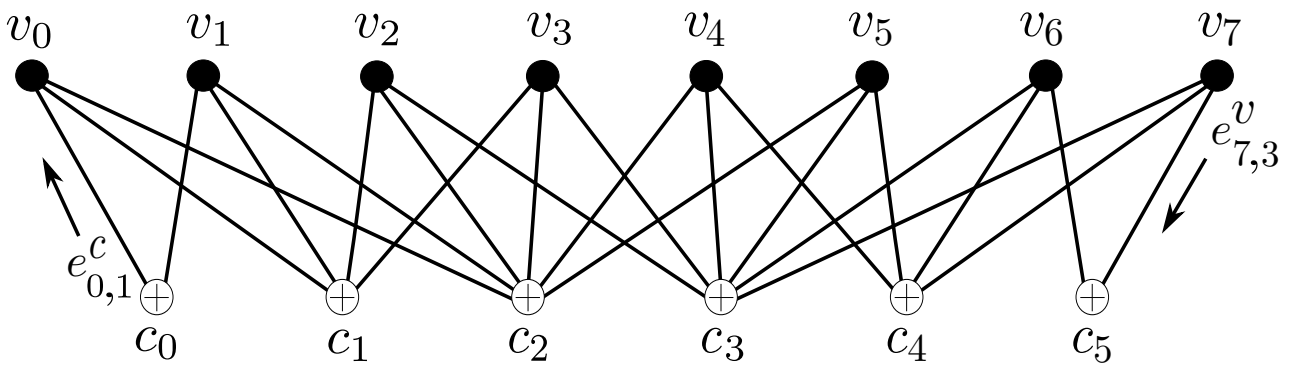

Fig. 1: An irregular protograph with $n_v = 8$ variable nodes and $n_c = 6$ check nodes.

This protograph is called irregular because the variable and check node degrees are not constant.

For the analysis of iterative decoding, it is useful to label the edges in $E$ from both a variable node and a check node perspective. Then $e^v_{y,l}$ indicates the $l$th edge emanating from variable node $v_y$. Similarly, $e^c_{x,m}$ denotes the $m$th edge emanating from check node $c_x$. Note that $l \in \{1, \ldots, \partial(v_y)\}$ and $m \in \{1, \ldots, \partial(c_x)\}$, where $\partial(v_y)$ and $\partial(c_x)$ denote the degree of variable node $v_y$ and check node $c_x$, respectively. It follows that if $e^v_{y,l}$ and $e^c_{x,m}$ define the same edge, $v_y$ is connected to $c_x$.

An ensemble of protograph-based LDPC block codes can be created from a base matrix $\mathbf{B}$ using a *copy-and-permute*

operation [3]. A parity-check matrix $\mathbf{H}$ from the ensemble of protograph-based LDPC block codes can then be obtained by replacing ones with an $N \times N$ permutation matrix and zeros with the $N \times N$ all zero matrix in the base matrix $\mathbf{B}$. In the case when a variable node and a check node are connected by $r$ repeated edges, the associated entry in $\mathbf{B}$ equals $r$ and the corresponding block in $\mathbf{H}$ consists of a summation of $r$ $N \times N$ permutation matrices. The *ensemble* is defined as the set of all possible parity-check matrices $\mathbf{H}$ that can be formed using this method.

By construction, every code in the resulting ensemble has the same node degrees and structure. The ensemble design rate is given as $R = 1 - n_c/n_v$. (In the case of puncturing, where $u \leq n_v$ is the number of variable nodes transmitted over the channel, $R = (n_v - n_c)/u$.) In addition, the sparsity condition of an LDPC matrix is satisfied for large $N$. The code created by applying the copy-and-permute operation to an $n_c \times n_v$ protograph base matrix $\mathbf{B}$ has block length $n = Nn_v$.

### A. Density evolution for protograph-based ensembles

Since every member of the protograph-based ensemble preserves the structure of the base protograph, density evolution analysis for the resulting codes can be performed within the protograph. We assume that belief propagation (BP) decoding is performed after transmission over a binary erasure channel (BEC) with erasure probability $\varepsilon$. In every decoding iteration, all of the check nodes are updated followed by all of the variable nodes. The messages that are passed between the nodes represent either an erasure or the correct symbol value (0 or 1).

Let $q^{(i)}(e^c_{x,m})$ denote the probability that the check to variable node message sent along edge $e^c_{x,m}$ in decoding iteration $i$ is an erasure. (Note that this will be the case if at least one of the incoming messages from other neighbouring variable nodes is erased.) Explicitly,

$$q^{(i)}\left(e^c_{x,m}\right) = 1 - \prod_{m' \neq m} \left(1 - p^{(i-1)}\left(e^c_{x,m'}\right)\right), \quad (1)$$

where $p^{(i-1)}(e^c_{x,m'})$ denotes the probability that the incoming message in the previous update of check node $x$ is an erasure, with $m,\ m' \in \{1, \ldots, \partial(c_x)\}$. In contrast, the variable to check node message sent along edge $e^v_{y,l}$ is an erasure if the incoming message from the channel and the messages from all the other neighbouring check nodes are erasures. This happens with probability $p^{(i)}(e^v_{y,l})$, where

$$p^{(i)}\left(e^v_{y,l}\right) = \varepsilon \prod_{l' \neq l} q^{(i)}\left(e^v_{y,l'}\right), \quad (2)$$

with $l,\ l' \in \{1, \ldots, \partial(v_y)\}$. The *density evolution threshold* of an ensemble is defined as the maximal value of the channel parameter $\varepsilon$ for which $p^{(i)}$ converges to zero as $i$ tends to infinity.

### B. Weight enumeration for protograph-based ensembles

The preserved structure of members of a protograph-based LDPC code ensemble also facilitates the calculation of average weight enumerators. An *ensemble average weight enumerator* $A_d$ tells us that, given a particular Hamming weight $d$, a typical member of the ensemble has $A_d$ codewords with Hamming weight $d$. Combinatorial techniques for calculating enumerators for protograph-based ensembles have been presented in [8] and [9].

The *asymptotic spectral shape function* of a code ensemble can be written as

$$r(\delta) = \limsup_{n \to \infty} \frac{\ln(A_d)}{n}, \quad (3)$$

where $\delta = d/n$, $d$ is the Hamming weight, $n$ is the block length, and $A_d$ is the ensemble average weight enumerator. Suppose the first positive zero crossing of $r(\delta)$ occurs at $\delta = \delta_{min}$. If $r(\delta)$ is negative in the range $0 < \delta < \delta_{min}$, then $\delta_{min}$ is called the *minimum distance growth rate* of the code ensemble, and we can say that the majority of codes in the ensemble have minimum distance $d \geq n\delta_{min}$.

## III. Terminated protograph-based LDPC convolutional codes

A rate $R = b/c$ (time-varying) binary LDPC convolutional code [1] can be defined as the set of infinite binary sequences $\mathbf{v}_{[-\infty,\infty]}$ that satisfy the equation $\mathbf{v}_{[-\infty,\infty]}\mathbf{H}^{\mathbf{T}}_{[-\infty,\infty]} = \mathbf{0}$, where

$$\mathbf{H}^T_{[-\infty,\infty]} = \begin{bmatrix} \ddots & & & \ddots & & \\ & \mathbf{H}^T_0(0) & \cdots & \mathbf{H}^T_{m_s}(m_s) & & \\ & & \ddots & & \ddots & \\ & & & \mathbf{H}^T_0(t) & \cdots & \mathbf{H}^T_{m_s}(t+m_s) \\ & & & \ddots & & \ddots \end{bmatrix}$$

is the transposed parity-check matrix, also called the *syndrome former matrix*. The binary $(c-b) \times c$ submatrices $\mathbf{H}_i(t)$, $i = 0, 1, \cdots, m_s$, satisfy the conditions that $\mathbf{H}_{m_s}(t) \neq \mathbf{0}$ for at least one $t \in \mathbb{Z}$ and that $\mathbf{H}_0(t)$ has full rank for all $t$. We call $m_s$ the *syndrome former memory* and $\nu_s = (m_s + 1) \cdot c$ the *decoding constraint length*. These parameters determine the width of the nonzero diagonal region of $\mathbf{H}_{[-\infty,\infty]}$. The sparsity of the parity-check matrix is insured by demanding that its rows have Hamming weight much less than $\nu_s$. The code is said to be regular if its parity-check matrix $\mathbf{H}_{[-\infty,\infty]}$ has exactly $J$ ones in every column and $K$ ones in every row. The code is irregular if its row and column weights are not constant, and the degree distribution is used to characterize the check and variable node degrees in the Tanner graph of the code. In general, the code is *time-varying*; a time-varying LDPC convolutional code is periodic with period $T$ if $\mathbf{H}_i(t)$ is periodic, i.e., $\mathbf{H}_i(t) = \mathbf{H}_i(t+T), \forall\ i, t$, and if $\mathbf{H}_i(t) = \mathbf{H}_i, \forall\ i, t$, the code is *time-invariant*.

### A. Constructing protograph-based LDPC convolutional codes

Analogously to block codes, an ensemble of LDPC convolutional codes can be constructed from a protograph. We proceed by forming a time-invariant infinite base matrix[1] with component $b_c \times b_v$ submatrices $\mathbf{B}_0, \mathbf{B}_1, \ldots, \mathbf{B}_{m_s}$ as follows:

$$\mathbf{B}_{[-\infty,\infty]} = \begin{bmatrix} \ddots & & \ddots & & & \\ & \mathbf{B}_{m_s} & \cdots & \mathbf{B}_0 & & \\ & & \ddots & & \ddots & \\ & & & \mathbf{B}_{m_s} & \cdots & \mathbf{B}_0 \\ & & & & \ddots & & \ddots \end{bmatrix}. \quad (4)$$

The infinite Tanner graph associated with $\mathbf{B}_{[-\infty,\infty]}$ can be regarded as a *convolutional protograph*. An ensemble of time-varying LDPC convolutional codes can be formed from $\mathbf{B}_{[-\infty,\infty]}$ using the protograph construction method based on $N \times N$ permutation matrices described in Section II.

### B. Forming terminated protograph-based LDPC convolutional codes

Suppose that we start the base matrix defined in $(4)$ at time $t = 0$ and terminate it after $L$ time instants. The resulting finite-length base matrix is given by

$$\mathbf{B}_{[0,L-1]} = \begin{bmatrix} \mathbf{B}_0 & & \\ \vdots & \ddots & \\ \mathbf{B}_{m_s} & & \mathbf{B}_0 \\ & \ddots & \vdots \\ & & \mathbf{B}_{m_s} \end{bmatrix}_{(L+m_s)b_c \times Lb_v}. \quad (5)$$

The matrix $\mathbf{B}_{[0,L-1]}$ can be considered as the base matrix of a terminated protograph-based LDPC convolutional code ensemble. Termination in this fashion results in a rate loss. Without puncturing, the design rate $R_L$ of the terminated code ensemble is equal to

$$R_L = 1 - \left(\frac{L+m_s}{L}\right)\frac{b_c}{b_v} = 1 - \left(\frac{L+m_s}{L}\right)(1-R), \quad (6)$$

where $R = 1 - Nb_c/Nb_v = 1 - b_c/b_v$ is the rate of the unterminated LDPC convolutional code ensemble. Note that, as the termination factor $L$ increases, the rate increases and approaches the rate of the unterminated LDPC convolutional code ensemble. The protograph-based LDPC block code ensemble associated with $\mathbf{B}_{[0,L-1]}$ can be studied using the analysis discussed in Section II.

[1] If the base matrix is binary, it represents the parity-check matrix of a rate $R = 1 - b_c/b_v$ time-invariant convolutional code with syndrome former memory $m_s$.

## IV. Analysis of terminated protograph-based LDPC convolutional codes

In this section, we begin by forming asymptotically regular LDPC block code ensembles by terminating several rate $R = 1/2$ protograph-based LDPC convolutional code ensembles with increasing complexity. The iterative decoding thresholds and minimum distance growth rates of the resulting LDPC block code ensembles are calculated and compared. We then show that the procedure can be applied to $(J, K)$-regular protograph-based LDPC convolutional codes with varying rates.

### A. An asymptotically regular $(3, 6)$ code family

Let $a = \gcd(J, K)$ denote the greatest common divisor of $J$ and $K$. Then there exist positive integers $J'$ and $K'$ such that $J = aJ'$ and $K = aK'$ with $\gcd(J', K') = 1$. It follows that the base matrix of a $(J, K)$-regular protograph-based LDPC convolutional code ensemble with syndrome former memory $m_s = a - 1$ can be defined as in (4), where the submatrices $\mathbf{B}_i$, $i = 0, \ldots, m_s$, are identical $J' \times K'$ matrices with all entries equal to one. (Note that, if $a = 1$, the syndrome former memory is equal to zero and the convolutional protograph is not fully connected.) For the $(3, 6)$-regular ensemble, we calculate $\gcd(J, K) = a = 3$ and the component submatrices of size $J' \times K' = b_c \times b_v = 1 \times 2$ are given as follows:

$$\mathbf{B}_0 = \begin{bmatrix} 1 & 1 \end{bmatrix} = \mathbf{B}_1 = \mathbf{B}_2.$$

Using these component submatrices, we can obtain the base matrix for a $(3, 6)$-regular LDPC convolutional code ensemble with syndrome former memory $m_s = 2$ as in (4).[2] Starting at time $t = 0$, the resulting terminated base matrix after $L$ time instants is

$$\mathbf{B}_{[0,L-1]} = \begin{bmatrix} \mathbf{B}_0 & & \\ \mathbf{B}_1 & \ddots & \\ \mathbf{B}_2 & & \\ & \ddots & \mathbf{B}_0 \\ & \ddots & \mathbf{B}_1 \\ & & \mathbf{B}_2 \end{bmatrix}_{(L+2) \times 2L}. \quad (7)$$

For $L \geq 3$, the ensemble design rate is

$$R_L = 1 - \frac{n_c}{n_v} = 1 - \frac{L+2}{2L} = \frac{L-2}{2L}.$$

Note that, while the terminated code ensembles approach the check node degree distribution of the $(3, 6)$-regular LDPC convolutional ensemble as $L \to \infty$, for finite $L$ the terminated ensembles have a reduced fraction of degree $6$ check nodes. For $L \geq 3$, the protograph has two degree $2$ check nodes, two degree $4$ check nodes, and $L - 2$ degree $6$ check nodes. By design, the variable node degree distribution remains constant for all $L$. The calculated minimum distance growth rates and BEC thresholds for these ensembles are given in Table I.

[2] This construction was presented as Example 1 in [7].

| $L$ | Rate $R_L$ | Growth rate $\delta_{min}^{(L)}$ | $\frac{\delta_{min}^{(L)} L}{m_s+1}$ | BEC threshold | Capacity $\varepsilon_{sh}$ | Gap to Capacity |
|---|---|---|---|---|---|---|
| 3 | 1/6 | 0.1419 | 0.142 | 0.714 | 0.833 | 0.119 |
| 4 | 1/4 | 0.0814 | 0.109 | 0.635 | 0.750 | 0.115 |
| 5 | 3/10 | 0.0573 | 0.096 | 0.588 | 0.700 | 0.112 |
| 6 | 1/3 | 0.0449 | 0.090 | 0.557 | 0.667 | 0.110 |
| 7 | 5/14 | 0.0374 | 0.087 | 0.537 | 0.643 | 0.106 |
| 8 | 3/8 | 0.0324 | 0.086 | 0.522 | 0.625 | 0.103 |
| 9 | 7/18 | 0.0287 | 0.086 | 0.512 | 0.611 | 0.099 |
| 10 | 2/5 | 0.0258 | 0.086 | 0.505 | 0.600 | 0.095 |
| 20 | 9/20 | 0.0129 | 0.086 | 0.488 | 0.550 | 0.062 |
| $\infty$ | 1/2 | 0 | | 0.488 | 0.500 | 0.012 |

TABLE I: Parameters for the terminated $(3,6)$-regular LDPC convolutional code ensembles.

As the termination factor $L$ tends to infinity, we observe that the minimum distance growth rate $\delta_{min}^{(L)}$ tends to zero.[3] This is consistent with similar results obtained for tail-biting LDPC convolutional code ensembles in [10]. We also observe from Table I that the scaled growth rates $\delta_{min}^{(L)} L/(m_s+1)$ converge to a fixed value as $L$ increases. A similar result was first observed in [11] for an ensemble of $(3,6)$-regular LDPC convolutional codes constructed from $N \times N$ permutation matrices, where it was shown that the scaled growth rates converged to a bound on the *free distance* growth rate of the unterminated LDPC convolutional code ensemble. This fact allows us to estimate the minimum distance growth rate $\delta_{min}^{(L)}$ for larger $L$, where the methods described in Section II-B become difficult to apply, by multiplying this bound on the free distance growth rate by $(m_s+1)/L$.

In addition to the convergence of the scaled minimum distance growth rate with increasing $L$, Table I also indicates that the BEC iterative decoding threshold converges to a constant value and that the gap to capacity decreases with increasing $L$. Since the distance growth rates decrease with $L$, this indicates the existence of a trade-off between distance growth rate and threshold. For this ensemble, the threshold approaches $\varepsilon^* = 0.488$ as $L \to \infty$. This is very close to the Shannon limit $\varepsilon_{sh} = 0.5$ for rate $R_\infty = 1/2$. Importantly, the threshold does not further decay as the termination factor $L$ increases. This remarkable result was first observed empirically in [12] for $(J,2J)$-regular ensembles constructed from $N \times N$ permutation matrices, and it was shown to be true for arbitrarily large $L$ in [13]. More recently, it has been shown in [14] that the iterative decoding thresholds of LDPC convolutional code ensembles on the BEC are equal to the optimal maximum a posteriori probability (MAP) decoding thresholds of their corresponding LDPC block code ensembles.

### B. *More asymptotically regular rate $R = 1/2$ code families*

Here, we consider how the thresholds and distance growth rates of other asymptotically regular rate $R = 1/2$ code families are affected by increasing the variable node degree $J$ to values greater than 3. Using component submatrices $\mathbf{B}_i = [\,1\;\;1\,], i = 0, \ldots, m_s = a = J-1$, (4) defines the convolutional base matrix of a rate $R = 1/2$, $(J,K)$-regular LDPC convolutional code ensemble. Terminating these codes using the procedure defined in Section III-B results in families of asymptotically regular $(J,K)$ LDPC block code ensembles. As we increase $J$, the complexity (measured as the average node degree) grows. Table II describes the complexity of the terminated ensembles with variable node degree $J$ and termination factor $L$.

| Aymptot. reg. ensemble | Rate $R_L$ | Variable node degree | Avg. check node degree |
|---|---|---|---|
| $(3,6)$ | $(L-2)/2L$ | 3 | $6L/(L+2)$ |
| $(4,8)$ | $(L-3)/2L$ | 4 | $8L/(L+3)$ |
| $(5,10)$ | $(L-4)/2L$ | 5 | $10L/(L+4)$ |
| $(J,2J)$ | $(L-J+1)/2L$ | $J$ | $2JL/(L+J-1)$ |

TABLE II: Complexity of the terminated rate $R = 1/2$ protograph-based LDPC convolutional code ensembles.

For finite $L$, the average check node degree of the asymptotically regular code ensemble is strictly less than that of the unterminated convolutional code ensemble. The check node degree increases with $L$, tending to the average check node degree of the unterminated ensemble as $L$ tends to infinity. The variable node degree remains constant at $J$ for all termination factors $L$.

Figure 2 plots the minimum distance growth rates for families of terminated code ensembles with $J = 3, 4$, and 5, some $(J,K)$-regular ensembles, and the Gilbert-Varshamov bound for the entire ensemble of block codes, where the values $J = 3, 4$, and 5 correspond to asymptotically regular $(3,6)$, $(4,8)$, and $(5,10)$ families, respectively.

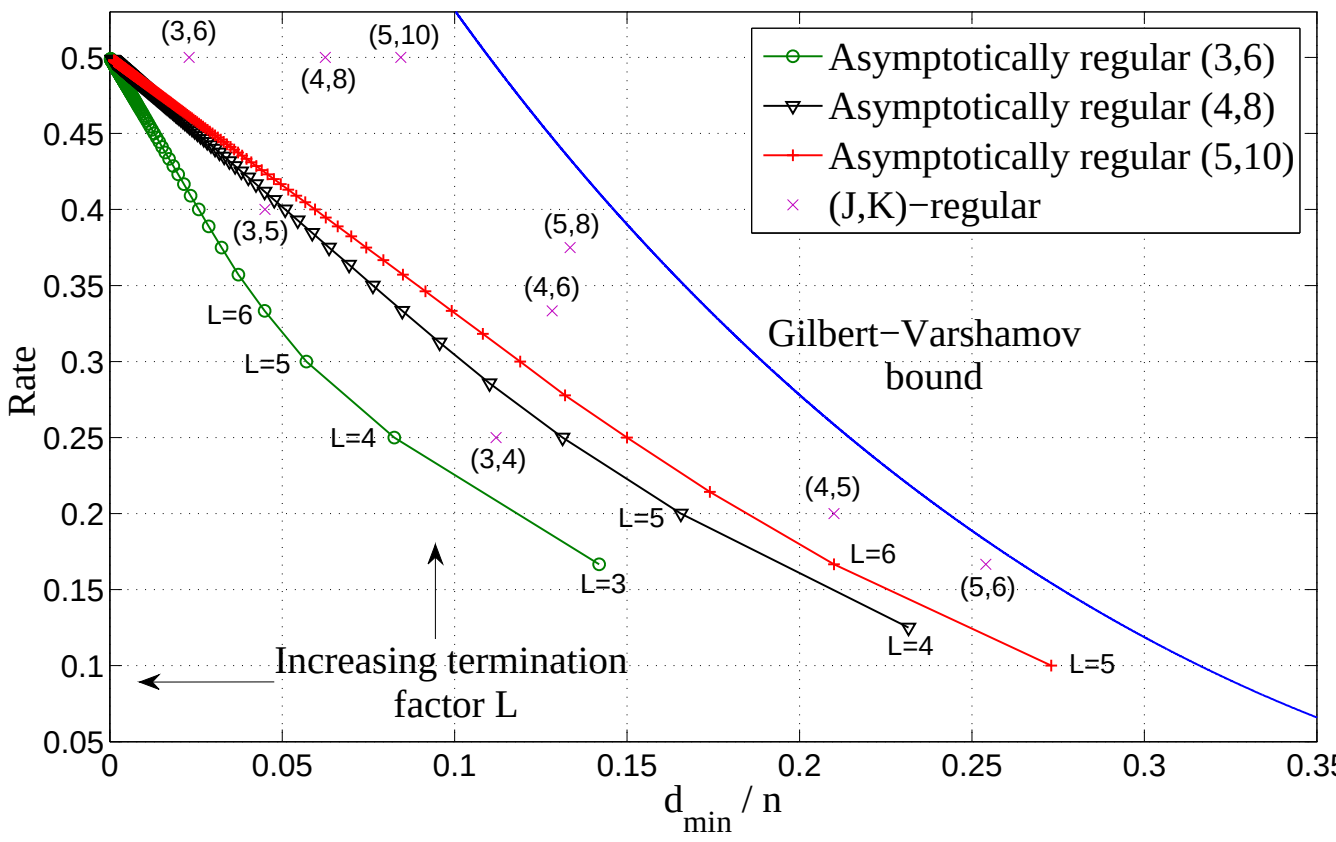

Fig. 2: Minimum distance growth rates for several families of terminated rate $R = 1/2$ protograph-based LDPC convolutional codes.

As with the asymptotically regular $(3,6)$ family analysed in Section IV-A, we find that increasing the termination factor $L$ results in declining minimum distance growth rates for the asymptotically regular $(4,8)$ and $(5,10)$ families. We again observe that the scaled minimum distance growth rates $\delta_{min}^{(L)} L/(m_s+1)$ converge as $L$ increases, which allows us to estimate the growth rates for $L \geq 10$. As expected, there is a significant increase observed for the growth rates of the $(4, 8)$ family compared to the $(3, 6)$ family. There is a further improvement for the asymptotically regular $(5, 10)$ family, but the increase is not as significant. We would expect this trend to continue as we further increase the variable node degree $J$.

[3] An infinite termination factor corresponds to the unterminated LDPC convolutional code ensemble. Using the techniques developed in [10], this convolutional code ensemble can be shown to be asymptotically good in the sense that the minimum *free distance* grows linearly with encoding constraint length.

Figure 3 plots the BEC iterative decoding thresholds for the asymptotically regular $(3, 6)$, $(4, 8)$, and $(5, 10)$ LDPC code families. We observe that the gap to capacity decreases as the termination factor $L$ increases. For regular ensembles of the same rate, one would expect the thresholds to worsen as we increase $J$ and $K$. Figure 3 shows that this is also the case for the asymptotically regular code families, for small termination factors $L$. Thus, by increasing $J$, and hence the complexity, we obtain a more pronounced trade-off between distance growth rates and threshold for small values of $L$. However, as the termination factor $L$ tends to infinity, we observe that the threshold of the asymptotically regular LDPC code families converge to a value close to capacity. This value improves as we increase $J$ ($\varepsilon^* = 0.4881$, $0.4977$, and $0.4994$ for the asymptotically regular $(3, 6)$, $(4, 8)$ and $(5, 10)$ LDPC code families, respectively). This indicates that, for large $L$, both the distance growth rates and the thresholds improve with increasing complexity. We would expect this trend to continue as we further increase the variable node degree $J$.

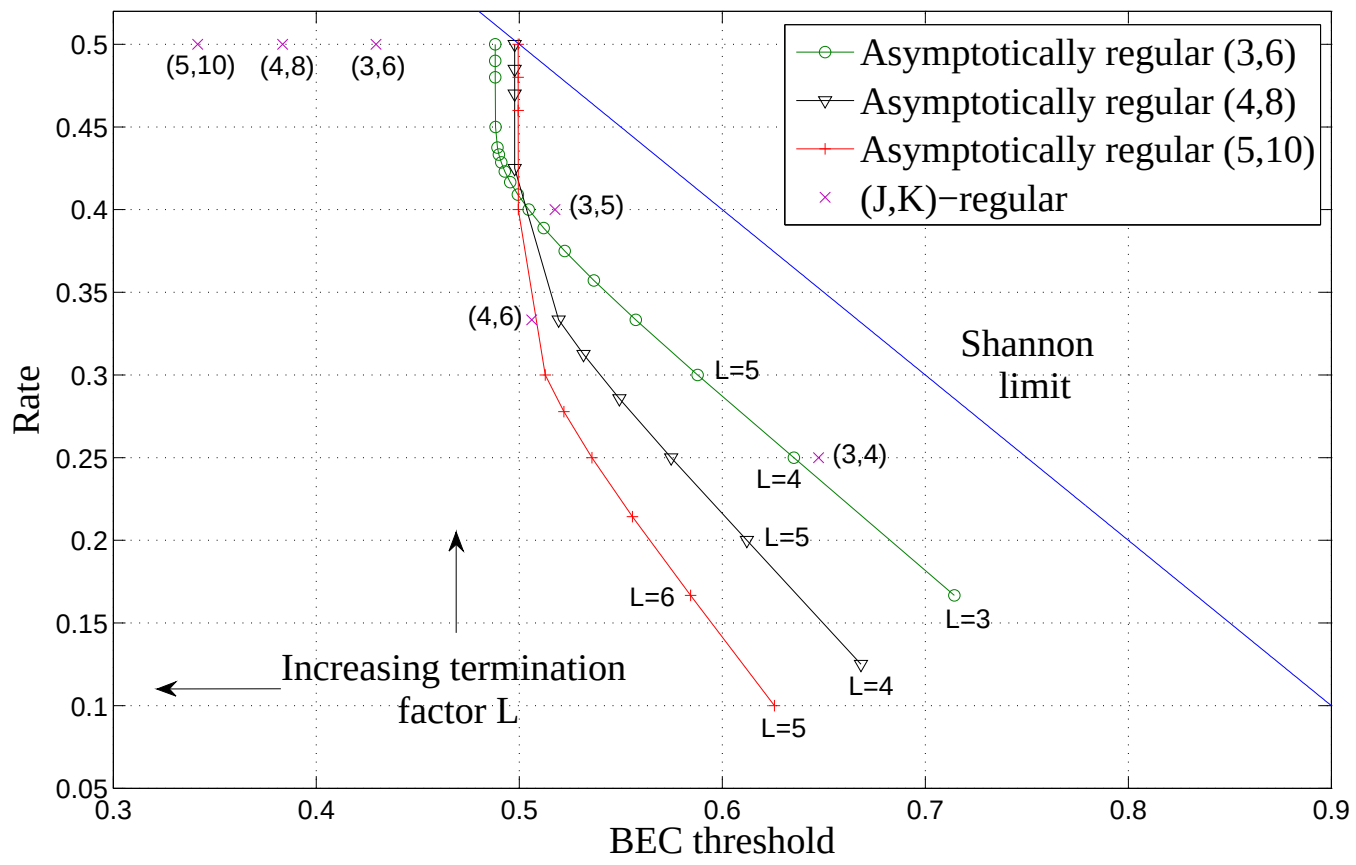

Fig. 3: BEC thresholds for several families of terminated rate $R = 1/2$ protograph-based LDPC convolutional codes.

### C. Other asymptotically regular code families

The procedure described in Section IV-A can be extended to form the base matrix of an arbitrary $(J, K)$-regular protograph-based LDPC convolutional code ensemble. For example, in the $(3, 9)$-regular case, $\gcd(3, 9) = 3 = a$, and the submatrices $\mathbf{B}_i$, $i = 0, \ldots, m_s = a - 1 = 2$, are identical $J' \times K' = 1 \times 3$ matrices with all entries equal to one. By using these component submatrices in (4), we obtain the base matrix for a $(3, 9)$-regular LDPC convolutional code ensemble with syndrome former memory $m_s = 2$. For termination factors $L \geq 2$, (5) defines the base matrix of a protograph-based LDPC code ensemble with two degree 3 check nodes, two degree 6 check nodes, and $L-2$ degree 9 check nodes; hence it is asymptotically regular. The rate of the asymptotically regular ensemble with termination factor $L \geq 2$ is $R_L = (2L-2)/3L$.

In the same way, we can construct the base matrices of $(3, 12)$- and $(4, 6)$-regular protograph-based LDPC convolutional code ensembles using the submatrices

$$\mathbf{B}_i = \left[\begin{array}{cccc} 1 & 1 & 1 & 1 \end{array}\right], i = 0, \ldots, m_s = a - 1 = 2, \text{ and}$$

$$\mathbf{B}_i = \left[\begin{array}{ccc} 1 & 1 & 1 \\ 1 & 1 & 1 \end{array}\right], i = 0, \ldots, m_s = a - 1 = 1,$$

respectively.

Figure 4 displays the BEC thresholds and growth rates for the asymptotically regular LDPC block code ensembles discussed in this section and some $(J, K)$-regular block code ensembles, along with the Shannon limit and the Gilbert-Varshamov bound, respectively. For each family, the iterative decoding threshold converges to a value close to the Shannon limit for $R_\infty$ as $L$ gets large. The design rates $R_L$ of the asymptotically regular ensembles, given by (6), cover a large range and approach the rate of the $(J, K)$-regular LDPC convolutional code ensemble. The range of achieveable code rate can be expanded by considering higher or lower rate $(J, K)$-regular convolutional code ensembles.

## V. Edge-spreading

As mentioned in Section IV-A, if $\gcd(J, K) = a = 1$ then $m_s = 0$ and the convolutional protograph is not fully connected. In other words, the base matrix (4) consists of disconnected blocks $\mathbf{B}_0$. This can be avoided by creating the submatrices $\mathbf{B}_0, \mathbf{B}_1, \ldots, \mathbf{B}_{m_s}$ using an *edge-spreading* technique [7]. Here, the edges of the protograph base matrix $\mathbf{B}$ are spread over the component submatrices such that $\mathbf{B}_0 + \mathbf{B}_1 + \ldots + \mathbf{B}_{m_s} = \mathbf{B}$. Note that the submatrices necessarily have the same size as $\mathbf{B}$ and the technique is not limited to the case $a = 1$, i.e., it can also be used when $\gcd(J, K) = a > 1$. In fact, as we note below in Example 4, the greatest common divisor method for forming component submatrices can be considered as a particular type of edge spreading. To illustrate the technique, we now compare the thresholds and distance growth rates of several families of asymptotically regular $(3, 6)$ LDPC code ensembles formed by edge spreading.

*Example* 1: Consider the following all-ones base matrix of size $n_c \times n_v = 3 \times 6$:

$$\mathbf{B} = \left[\begin{array}{cccccc} 1 & 1 & 1 & 1 & 1 & 1 \\ 1 & 1 & 1 & 1 & 1 & 1 \\ 1 & 1 & 1 & 1 & 1 & 1 \end{array}\right]. \tag{8}$$

Using the component submatrices,

$$\mathbf{B}_0 = \left[\begin{array}{cccccc} 1 & 1 & 0 & 0 & 0 & 0 \\ 0 & 0 & 1 & 1 & 0 & 0 \\ 0 & 0 & 0 & 0 & 1 & 1 \end{array}\right], \mathbf{B}_1 = \left[\begin{array}{cccccc} 0 & 0 & 1 & 1 & 1 & 1 \\ 1 & 1 & 0 & 0 & 1 & 1 \\ 1 & 1 & 1 & 1 & 0 & 0 \end{array}\right],$$

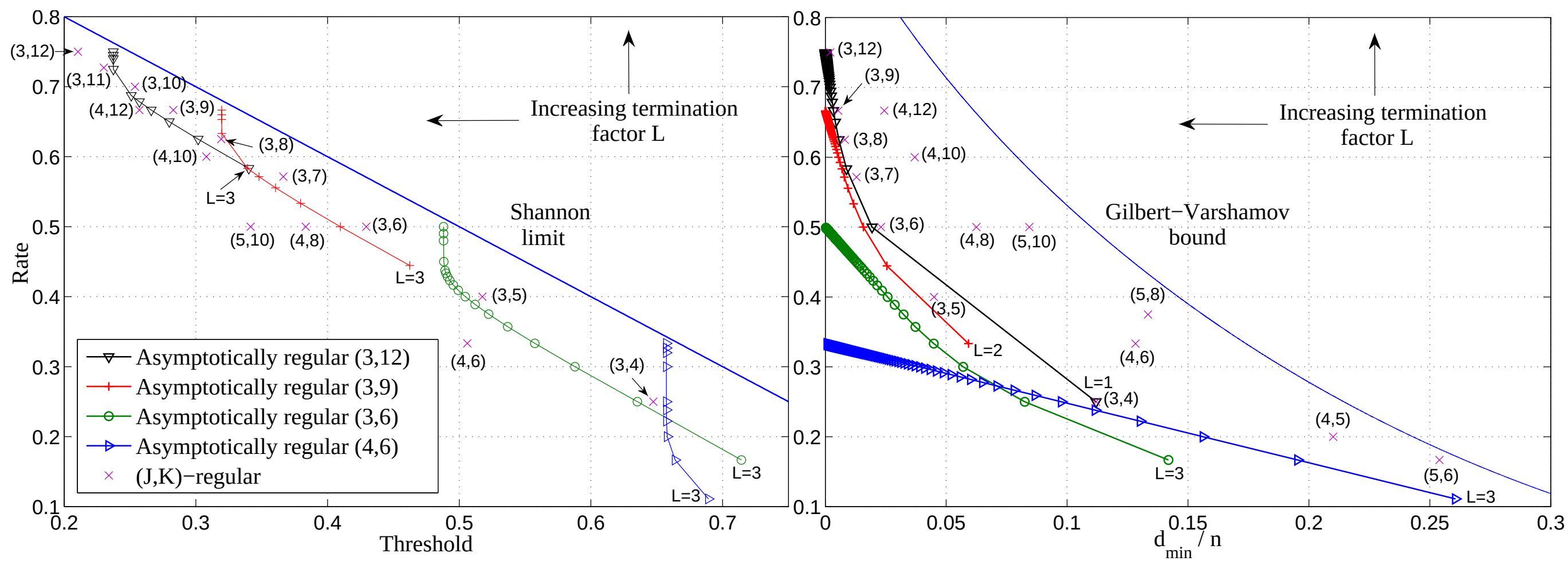

Fig. 4: BEC thresholds and minimum distance growth rates for families of asymptotically regular $(J, K)$ LDPC convolutional code ensembles.

we can form the base matrix of a $(3,6)$-regular protograph-based LDPC convolutional code with syndrome former memory $m_s = 1$ as in (4). Note that $\mathbf{B}_0 + \mathbf{B}_1 = \mathbf{B}$. Figure 5 shows the associated convolutional protograph obtained using component submatrices $\mathbf{B}_0$ and $\mathbf{B}_1$, along with the termination factors $L$ that form asymptotically regular ensembles.

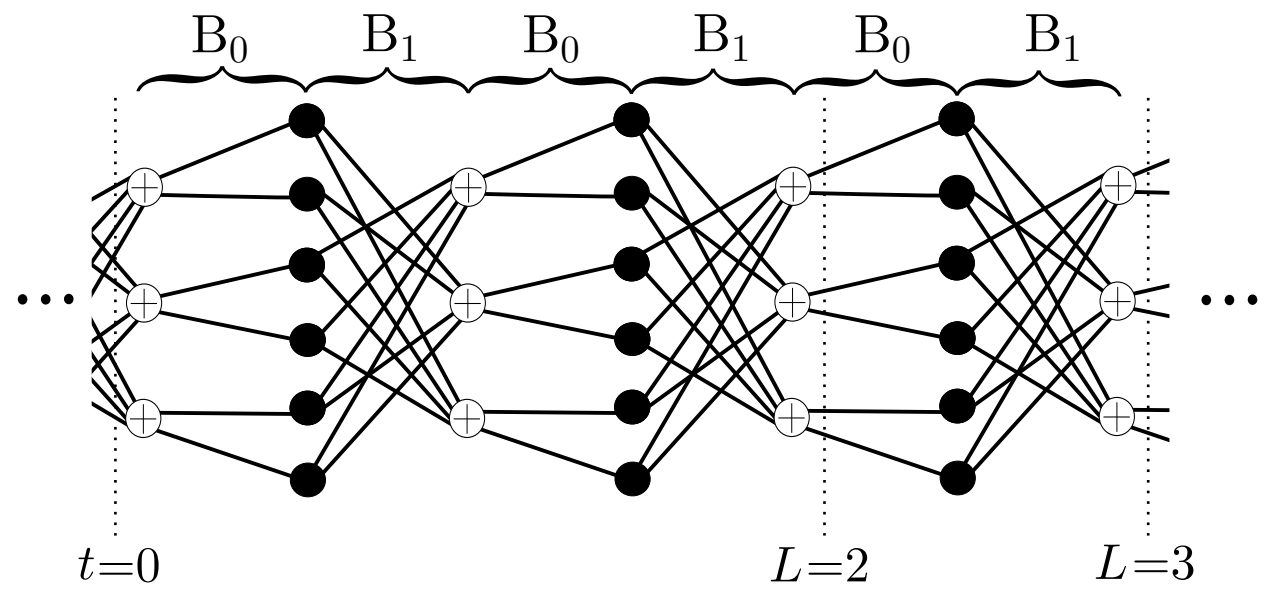

Fig. 5: The convolutional protograph of Example 1, along with some termination factors for increasing $L$.

The resulting design rate of the terminated code ensemble is $R_L = (L-1)/2L$. The terminated protograph has three degree 2 check nodes, three degree 4 check nodes, and $3L-3$ degree 6 checks, so it is an asymptotically regular $(3,6)$ ensemble.

*Example* 2: The following component submatrices of (8) have only degree 3 check nodes:

$$\mathbf{B}_0 = \begin{bmatrix} 1 & 1 & 1 & 0 & 0 & 0 \\ 0 & 1 & 1 & 1 & 0 & 0 \\ 0 & 0 & 0 & 1 & 1 & 1 \end{bmatrix} \text{ and } \mathbf{B}_1 = \mathbf{B} - \mathbf{B}_0.$$

Using $\mathbf{B}_0$ and $\mathbf{B}_1$ as given above, the asymptotically regular $(3,6)$ ensemble defined by (5) has six degree 3 check nodes and $3L-3$ degree 6 checks for termination factors $L \geq 2$. The protographs in this terminated family will be highly regular with no degree 2 check nodes.

*Example* 3: In order to reduce the memory requirements for implementing the codes, it is also interesting to consider repeated edges. Consider the following $n_c \times n_v = 1 \times 2$ base matrix

$$\mathbf{B} = \begin{bmatrix} 3 & 3 \end{bmatrix}.$$

The edges of $\mathbf{B}$ can be spread as follows:

$$\mathbf{B}_0 = \begin{bmatrix} 2 & 1 \end{bmatrix} \text{ and } \mathbf{B}_1 = \mathbf{B} - \mathbf{B}_0 = \begin{bmatrix} 1 & 2 \end{bmatrix}.$$

Using these component submatrices, the base matrix (4) defines a $(3,6)$-regular protograph-based LDPC convolutional code. Figure 6 shows the associated convolutional protograph, along with the termination factors that form asymptotically regular ensembles. As with Examples 1 and 2, this ensemble has syndrome former memory $m_s = 1$; however, the decoding constraint length is $\nu_s = (m_s+1)b_v = (m_s+1)n_v = 4N$, whereas $\nu_s = 12N$ for Examples 1 and 2.

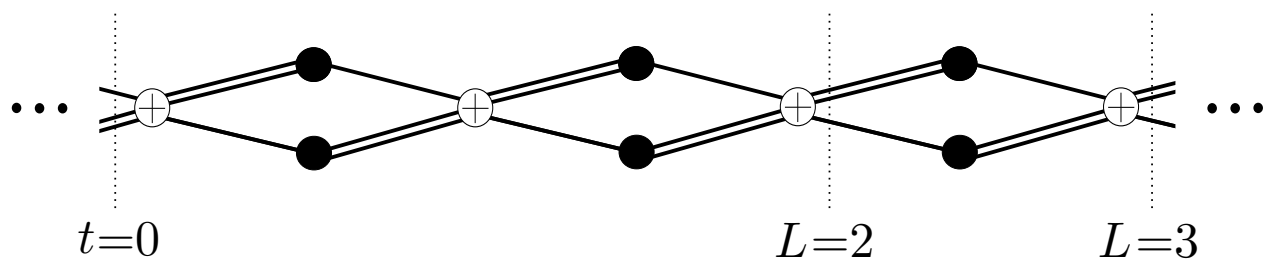

Fig. 6: The convolutional protograph of Example 3, along with some termination factors for increasing $L$.

*Example* 4: Using edge spreading, it is also possible to form base matrices that define the same $(J,K)$-regular LDPC convolutional code ensembles as the examples discussed in Section IV, which were formed by using a method based on the greatest common divisor of $J$ and $K$. The greatest common divisor method is equivalent to the particular edge spreading of a $J' \times K'$ base matrix $\mathbf{B}$ with all entries equal to $a$ into $m_s + 1 = a$ all-one $J' \times K'$ component submatrices. For example, for the $(3,6)$-regular ensemble, gcd$(J,K) = a = 3$, $J' = 1$, and $K' = 2$. Then, by splitting $\mathbf{B} = \begin{bmatrix} 3 & 3 \end{bmatrix}$ into $\mathbf{B}_0 = \mathbf{B}_1 = \mathbf{B}_2 = \begin{bmatrix} 1 & 1 \end{bmatrix}$, with $\mathbf{B}_0 + \mathbf{B}_1 + \mathbf{B}_2 = \mathbf{B}$, we have the same component submatrices as the asymptotically regular $(3,6)$ family presented in Section IV.

| $L$ | Rate | Example 1 | | Example 2 | | Example 3 | | Example 4 | |
|---|---|---|---|---|---|---|---|---|---|
| | $R_L$ | $\varepsilon^*$ | $\delta_{min}^{(L)}$ | $\varepsilon^*$ | $\delta_{min}^{(L)}$ | $\varepsilon^*$ | $\delta_{min}^{(L)}$ | $\varepsilon^*$ | $\delta_{min}^{(L)}$ |
| 2 | 1/4 | 0.6358 | 0.0873 | 0.6471 | 0.0920 | 0.6448 | 0.0950 | 0.6353 | 0.0814 |
| 3 | 1/3 | 0.5600 | 0.0496 | 0.5673 | 0.0511 | 0.5671 | 0.0524 | 0.5574 | 0.0449 |
| 4 | 3/8 | 0.5249 | 0.0362 | 0.5298 | 0.0367 | 0.5301 | 0.0375 | 0.5223 | 0.0324 |
| 5 | 2/5 | 0.5064 | 0.0289 | 0.5098 | 0.0291 | 0.5103 | 0.0298 | 0.5046 | 0.0258 |
| 6 | 5/12 | 0.4965 | 0.0241 | 0.4989 | 0.0243 | 0.4993 | 0.0248 | 0.4955 | 0.0215 |
| 7 | 3/7 | 0.4914 | 0.0206 | 0.4930 | 0.0208 | 0.4933 | 0.0213 | 0.4911 | 0.0184 |
| 8 | 7/16 | 0.4893 | 0.0180 | 0.4902 | 0.0182 | 0.4903 | 0.0186 | 0.4892 | 0.0161 |
| 20 | 19/40 | 0.4881 | 0.0072 | 0.4881 | 0.0072 | 0.4881 | 0.0074 | 0.4881 | 0.0065 |
| $\infty$ | 1/2 | 0.4881 | 0 | 0.4881 | 0 | 0.4881 | 0 | 0.4881 | 0 |

TABLE III: BEC thresholds and distance growth rates for various asymptotically regular $(3,6)$ LDPC code families constructed by edge spreading. The rate of these families is given as $R = (L-1)/2L$.

Moreover, the same convolutional base matrix can be formed from different edge spreadings. For example, consider the all-ones base matrix $\mathbf{B}$ of size $3 \times 6$. Using edge spreading, we can form the following component submatrices:

$$\mathbf{B}_0 = \begin{bmatrix} 1 & 1 & 0 & 0 & 0 & 0 \\ 1 & 1 & 1 & 1 & 0 & 0 \\ 1 & 1 & 1 & 1 & 1 & 1 \end{bmatrix} \text{ and } \mathbf{B}_1 = \mathbf{B} - \mathbf{B}_0.$$

Using these component submatrices in (4), we obtain the base matrix of a rate $R = 3/6$, $(3,6)$-regular LDPC convolutional code ensemble with syndrome former memory $m_s = 1$. This convolutional base matrix is identical to the base matrix that was constructed in Section IV-A using the greatest common divisor method, which resulted in an equivalent rate $R = 1/2$, $(3,6)$-regular LDPC convolutional code ensemble with $m_s = 2$. In this example, we use the $m_s = 1$ interpretation of the base matrix to form an asymptotically regular $(3,6)$ family so that we have equivalent rates $R_L$ for comparison with the other examples. The terminated base matrices $\mathbf{B}_{[0,L-1]}$ here can be obtained using termination factors $L = 2k$ in (7) for $k = 2, 3, \ldots$. Note that, in this sense, the asymptotically regular family defined using the greatest common divisor method is more flexible, since it can achieve ensembles with more finely grained design rates than those constructed in this example.

The thresholds and distance growth rates calculated for the asymptotically regular $(3,6)$ ensembles of Examples 1-4 are displayed in Table III. An interesting observation is that by eliminating the degree 2 check nodes, Examples 2 and 3 display larger growth rates and better thresholds than Examples 1 and 4. All of the thresholds converge to the same value $\varepsilon^* = 0.4881$ as $L \to \infty$, which is equal to the optimal MAP decoding threshold of $(3,6)$-regular ensembles. Example 3, which has the smallest decoding constraint length, achieves the best distance growth rates. This can most likely be attributed to having a larger proportion of non-zero elements in $\mathbf{B}_{[0,L-1]}$, i.e., a denser base matrix. For termination factors $L = 2$ and 3, Example 2 has the best thresholds, but for $L \geq 4$ the repeated edge Example 3 has both the best growth rates and thresholds.[4]

There are many ways of spreading the edges among the component submatrices of a base matrix $\mathbf{B}$, and different constructions can result in varying thresholds and ensemble growth rates. Choices containing all-zero rows and/or columns in the submatrices should be avoided, since they can lead to disconnected subgraphs. Note that simple row and column permutations (applied to all component submatrices simultaneously) do not affect the graph structure, and so, in turn, they do not affect the threshold and distance growth rate of the ensemble. A good threshold value is expected when the checks at time $t = 0$ have low degree (but at least degree 2).

[4]Constructing LDPC convolutional code ensembles from protographs with repeated edges in order to reduce memory requirements has recently been shown to improve the performance of a windowed decoder [15].

## VI. CONCLUSIONS

We have provided a construction technique for families of asymptotically regular LDPC block code ensembles formed by terminating $(J,K)$-regular protograph-based LDPC convolutional codes. By varying the termination length, we obtain a large selection of LDPC block code ensembles with varying code rates and substantially better iterative decoding thresholds than those of $(J,K)$-regular LDPC block code ensembles, despite the fact that the terminated ensembles are almost regular. By means of an asymptotic weight enumerator analysis, we showed that the minimum distance grows linearly with block length for all of the ensembles in these families, i.e., the ensembles are asymptotically good. As the termination length increases, we obtain a family of codes with capacity approaching iterative decoding thresholds and declining minimum distance growth rates.

It was also shown that, by increasing the complexity of the component submatrices forming the LDPC convolutional code ensemble, the minimum distance growth rates can be improved while maintaining a capacity approaching threshold. Further, using an edge spreading technique, we showed that both the iterative decoding threshold and the minimum distance growth rate of the ensemble can be improved by carefully choosing the component submatrices. As a result of the variable node degree design, we insure fast convergence rates and thresholds close to capacity. The discussion in this paper was limited to the BEC; however, based on the results of [13], we expect to observe similar behaviour for the additive white Gaussian noise channel. In practice, the design parameter $L$ adds an additional degree of freedom to existing block code designs. Starting from any LDPC block code, it is possible to derive terminated convolutional codes that share the same encoding and decoding architecture for arbitrary $L$.

## Acknowledgements

This work was partially supported by NSF Grant CCF08-30650 and NASA Grant NNX07AK53G.